\documentstyle[twocolumn,seceq,letter,epsf]{jpsj}
\def\runtitle{Numerical Renormalization Group Study of Kondo Effect
in Unconventional Superconductors}
\def\runauthor{Masashige {\sc Matsumoto} and Mikito {\sc Koga}}

\title{
Numerical Renormalization Group Study of Kondo Effect
in Unconventional Superconductors
}

\author{
Masashige {\sc Matsumoto} and Mikito {\sc Koga}$^1$
}

\inst{
Department of Physics, Faculty of Science, Shizuoka University, 836 Oya,
Shizuoka 422-8529, Japan \\
$^1$Department of Physics, Faculty of Education, Shizuoka University, 836 Oya,
Shizuoka 422-8529, Japan
}

\recdate
{June 28, 2001}

\abst{
Orbital degrees of freedom of a Cooper pair
play an important role in the unconventional superconductivity.
To elucidate the orbital effect in the Kondo problem,
we investigated a single magnetic impurity coupled to Cooper pairs
with a $p_x +i p_y$ ($d_{x^2-y^2}+id_{xy}$) symmetry
using the numerical renormalization group method.
It is found that the ground state is always a spin doublet.
The analytical solution for the strong coupling limit explicitly shows
that the orbital dynamics of the Cooper pair generates the spin 1/2 of the ground state.
}

\kword{
Kondo effect, unconventional superconductivity, numerical renormalization group
}
\begin{document}
\sloppy
\maketitle
\renewcommand{\theequation}{\arabic{equation}}
\newcommand{\ri}{{\rm i}}

Unconventional superconductivity is characterized by an angular momentum
($p$, $d$, or $f$) of its Cooper pair.
It is very sensitive to impurities and surface boundaries, differing from the
standard BCS ($s$-wave) superconductivity.
As a result, it displays various phenomena associated with
breaking of the Cooper pairs.
In the last decade, much attention has been paid to the effects of
non-magnetic impurities and static boundaries.
\cite{Matsumoto95-1,Salkola96,Pan,Hu,Matsumoto95-2,Covington,Fogelstrom}
At present, the attention shifts to the study of magnetic impurities
in the unconventional superconductivity.
Recent experimental studies provided evidence of induced local moments
with spin $S = 1/2$ in Zn-doped $d$-wave superconductors.
\cite{Alloul,Mahajan,Sidis}
A possibility of the Kondo effect was indicated in impurity-doped YBa$_2$Cu$_3$O$_{7-\delta}$.
\cite{Sisson,Bobroff}
At the present stage, the search for the Kondo behavior in unconventional superconductors
is still in progress.

A local spin in a superconducting state causes the following significant effects:
(1) breaking time-reversal invariance and (2) dynamical coupling with quasiparticles
due to the Kondo effect.
For the first effect, the local spin can be treated in a classical way.
\cite{Salkola97}
On the other hand, many-body correlations play an important role in causing the second effect.
In the case of $s$-wave superconductors, we have a competition between the Kondo effect
and the superconducting energy gap $\Delta$.
Even if the gap exists in the density of quasiparticle states, the local spin forms a spin
singlet with the quasiparticles under the condition $T_{\rm K} > \Delta$, where $T_{\rm K}$ is
the Kondo temperature.
\cite{Satori,Sakai}
A similar competition is found in a gapless $d_{x^2-y^2}$-wave superconductor
\cite{Simon}.
In this case, the gapless density of states stabilizes the Kondo singlet.

In this letter, we study a local spin coupled to a gapped superconducting state
with a $p_x + i p_y$ (or $d_{x^2 - y^2} + i d_{xy}$) symmetry.
The $d$-vector for the $p_x +i p_y$-wave
is given by $\mbox{\boldmath$d$}_k=(k_x +i k_y)\hat{z}$,
where $\hat{z}$ is a unit vector in the $z$ axis.
Recently, the property of the $p_x + i p_y$-wave superconductivity
has been investigated extensively,
which was motivated by the discovery of the spin triplet superconductor Sr$_2$RuO$_4$.
\cite{Maeno,Rice,Luke,Ishida,Sigrist}
It is analogous to the property of the A phase of the superfluid $^3$He.
In recent theoretical studies of the $p_x + i p_y$ superconductivity,
unique physics concepts such as vortex charging effect,
\cite{Goryo00,Matsumoto00-1}
spontaneous current
\cite{Matsumoto99-1,Okuno}
and spontaneous Hall effect have been reported.
\cite{Goryo99,Furusaki}
The effect of magnetic impurities should also reveal rich properties
peculiar to the orbital degrees of freedom of the Cooper pair.

The Cooper pair of the unconventional superconductivity
is constructed by conduction electrons with various angular momenta.
In the $p_x +i p_y$-wave state,
only two angular momenta are coupled to each other.
It enables us to apply the numerical renormalization group (NRG) method.
\cite{Wilson}
This is the first NRG study of such a non-$s$-wave superconducting state.
Our result shows that the coupled conduction electron system
displays a new type of Kondo effect.
We find that the ground state is a spin doublet
in all the parameter region of $T_{\rm K} / \Delta$.
The spin singlet ground state cannot be realized
even in the strong coupling region ($T_{\rm K}/\Delta \gg 1$),
since the paired electrons generate the spin 1/2 of the ground state.
It is different from the $s$-wave case
where the interchange of spin doublet and singlet ground states occurs.
\cite{Satori}
We discuss the difference between the $s$-wave and $p_x +i p_y$-wave,
and show how the spin doublet ground state is realized for the $p_x + i p_y$-wave
in the strong coupling limit.

In this letter, we study a single magnetic impurity
at the center in two-dimensional superconducting systems.
Since we treat short-range scattering here,
the impurity couples with only the electrons having no angular momentum ($l=0$).
For the $p_x +i p_y$-wave, the total angular momentum of the Cooper pair is equal to one.
The order parameter is expressed by $\Delta e^{i\phi_k}$,
where $\phi_k$ is the angle of the Fermi vector measured from the $k_x$ axis.
Since the angular momentum is a good quantum number for the $p_x +i p_y$-wave,
the $l=0$ and $l=1$ orbitals are coupled,
and they are decoupled from the other orbitals.
This is the reason why the Kondo problem in the $p_x +i p_y$-wave state
can be treated within the two angular momentum spaces.
Therefore, we can apply the NRG method to the Kondo problem
for the $p_x +i p_y$-wave as we do to the two-channel Kondo problem.
\cite{Pang}
Let us begin with the following Hamiltonian $H=H_{p_x +i p_y}+H_{\rm imp}$:
\begin{eqnarray}
&&H_{p_x +i p_y}
= \sum_{k\sigma} \sum_{l=0,1} \epsilon_k a_{kl\sigma}^\dagger a_{kl\sigma} \cr
&&~~~~~~~~~~~+ \sum_{k\sigma}
     \Bigl[
          i\Delta a_{k1\sigma}^\dagger a_{k0,-\sigma}^\dagger + {\rm H.c.}
     \Bigr], \cr
&&H_{\rm imp} = \frac{1}{2} \sum_{kk'\sigma\sigma'}
   ( -J \mbox{\boldmath$S$} \cdot \mbox{\boldmath$\sigma$}_{\sigma\sigma'} 
     +V \delta_{\sigma,\sigma'} ) a_{k0\sigma}^\dagger a_{k'0\sigma'}.
\label{eqn:Hamiltonian}
\end{eqnarray}
Here, $H_{p_x +i p_y}$ is the Hamiltonian for the $p_x +i p_y$-wave pairing state,
and $H_{\rm imp}$ represents exchange interaction and potential scattering of conduction electrons
due to the impurity.
$\epsilon_k$ is the kinetic energy of the conduction electron with wave number $k$,
and $\Delta$ is the $p_x +i p_y$-wave order parameter.
The subscripts $l$(=0,1) and $\sigma(=\uparrow,\downarrow)$ in $a_{kl\sigma}$ represent
the angular momentum and spin of the conduction electron, respectively.
$\mbox{\boldmath$S$}$ and $\mbox{\boldmath$\sigma$}$ are the $S=1/2$ spin operator
and the Pauli matrix for the magnetic impurity and conduction electrons, respectively.
$J(<0)$ and $V$ are the antiferromagnetic and non-magnetic couplings, respectively.
Note that the model given by eq. (\ref{eqn:Hamiltonian}) is not a simple two-band Kondo model.
Only the $l=0$ conduction electrons are coupled directly to the local spin.
In the normal state,
the $l=1$ electrons are completely decoupled from the local spin.
However, in the superconducting state,
they are coupled to the $l=0$ conduction electrons via the $p_x +i p_y$-wave order parameter.
We have to take the $l=1$ electrons into account as well for the $p_x +i p_y$-wave.
We can elucidate the orbital effect of the $p_x +i p_y$-wave and compare it with the $s$-wave,
since both superconductors are full-gap systems
and have the same density of states.
We can obtain a similar Hamiltonian for a $d_{x^2-y^2}+id_{xy}$-wave
if we replace $i\Delta$ by $-\sigma\Delta$ and $l=1$ by $l=2$.

In accordance with the Wilson's NRG procedure,
\cite{Wilson}
the bare conduction band is discretized logarithmically
and the Hamiltonian (\ref{eqn:Hamiltonian}) can be transformed into the following hopping type
with staggered potentials
\cite{Satori,Sakai}
for both $p_x +i p_y$-wave and $d_{x^2-y^2}+id_{xy}$-wave pairing states:
\begin{eqnarray}
&&H_{N+1} = \Lambda^{1/2}H_N + \sum_{\tau\sigma}
  \Bigl[
    \varepsilon_N ( c_{N+1,\tau\sigma}^\dagger c_{N\tau\sigma} + {\rm H.c.} ) \cr
&&~~~~~~~~~~~~~~~~~~~~~
     +(-1)^N\Lambda^{N/2}\tau\tilde{\Delta}
        c_{N+1,\tau\sigma}^\dagger c_{N+1,\tau\sigma}
  \Bigr], \cr
&&H_0 = \Lambda^{-1/2}
  \Bigl[ \frac{1}{2}
    \sum_{\tau\tau'\sigma\sigma'}
      ( - \tilde{J} \mbox{\boldmath$S$} \cdot \mbox{\boldmath$\sigma$}_{\sigma\sigma'}
        + \tilde{V} \delta_{\sigma,\sigma'} )
       c_{0\tau\sigma}^\dagger c_{0\tau'\sigma'} \cr
&&~~~~~~~~~~~~~~~ -\sum_{\tau\sigma} \tau\tilde{\Delta} c_{0\tau\sigma}^\dagger c_{0\tau\sigma}
  \Bigr].
\label{eqn:NRG}
\end{eqnarray}
The subscript $\tau=\pm$ represents the two channels constructed by the $l=0$ and $l=1$ orbitals.
Note that the channels are not independent
due to the channel-flip ($\sum_{\tau\tau'}$) terms.
This means that only one of the paired electrons ($l=0$)
couples with the local spin as in eq. (\ref{eqn:Hamiltonian}),
since the operator of the conduction electron at the impurity is written by
\begin{equation}
\sum_k a_{k,l=0,\sigma} = c_{N=0,\tau=+,\sigma} + c_{N=0,\tau=-,\sigma}.
\end{equation}
Here, $c_{N\tau\sigma}$ is the operator of the NRG fermion quasiparticle in the $N$-th shell.
In eq. (\ref{eqn:NRG}), $\varepsilon_N$ is given by
\begin{equation}
\varepsilon_N = [1-\Lambda^{-(N+1)}][1-\Lambda^{-(2N+1)}]^{-1/2}[1-\Lambda^{-(2N+3)}]^{-1/2},
\end{equation}
where $\Lambda$ is the discretization parameter.
The values with tilde are normalized by $(1+\Lambda^{-1})/2$.
For simplicity, we used the same values
for both the superconducting cutoff energy and the band width,
which does not alter the results.
\cite{Sakai}
Throughout this letter,
we keep the lowest-lying 500 states at each renormalization step and take $\Lambda=3$.

\begin{figure}[t]
\begin{minipage}{8cm}
\epsfxsize=8cm
\epsfbox{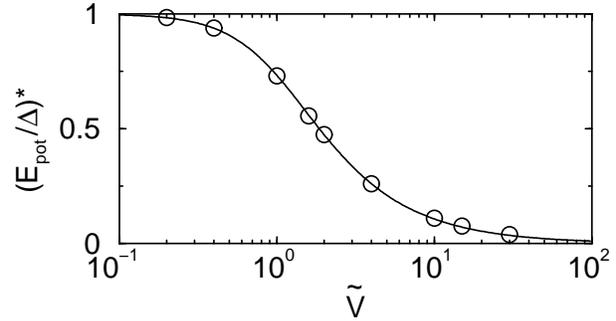}
\end{minipage}
\caption{
Bound state (one-hole excitation) energy $(E_{\rm pot}/\Delta)^*$
measured from the singlet ground state energy for each $\tilde{V}$.
The parameters are fixed at $\tilde{J}=0$ and $\tilde{\Delta}=0.01$.
The circles are the NRG results.
The solid line represents $[1+(\alpha \tilde{V})^2]^{-1/2}$,
where $\alpha=0.929 $ is used to fit the NRG results.
The $\tilde{V}$ dependence of $(E_{\rm pot}/\Delta)^*$ is independent of $\tilde{\Delta}$.
}
\label{fig:1}
\end{figure}
First, we discuss the case of $\tilde{J}=0$ to demonstrate the reliability of our NRG results.
Figure \ref{fig:1} shows a finite $\tilde{V}$ case
in which an excited state (bound state) appears below the gap.
The appearance of the bound state is due to the pair breaking effect
of the potential scattering (non-magnetic impurity effect).
The ratio of the renormalized bound state energy and the renormalized energy gap
converges as the renormalization step $N$ increases.
The convergence value is represented by $(E_{\rm pot}/\Delta)^*$.
In Fig. \ref{fig:1}, $(E_{\rm pot}/\Delta)^*$ decreases with $\tilde{V}$
as expected.
The NRG result is in good agreement with the analytic solution
given by a function of $\tilde{V}$:
\cite{Okuno}
\begin{equation}
E_{\rm pot} = \frac{\Delta}{\sqrt{1+(\alpha\tilde{V})^2}}.
\end{equation}

Let us consider the $\tilde{J}\neq 0$ case.
Since we have confirmed that the potential term $\tilde{V}$ does not change the ground state,
we restrict ourselves to the $\tilde{V}=0$ case.
When $\tilde{J}\rightarrow 0$ and $\tilde{\Delta}\neq 0$,
the impurity spin is decoupled from the quasiparticles
and there is no bound state below the superconducting energy gap as expected.
In this case, the NRG Hamiltonian has two independent $\tau=\pm$ channels
which have the same form as the $s$-wave except for the sign of the order parameter
depending on $\tau$.
The NRG energy level structure for the $p_x +i p_y$-wave has particle-hole symmetry.
In the opposite limit ($\tilde{J}\neq 0$ and $\tilde{\Delta}=0$),
the NRG energy level structure at the fixed point
is given by the admixture of a strong coupling type and a free spin type.
The former is for the $l=0$ orbital and the latter is for the $l=1$.
The NRG energy spectra can be described by filling quasiparticles
in one-particle energy levels.
In the $\tilde{\Delta}=0$ case, an energy level lies just on the Fermi energy.
Since this level can be occupied by at most two particles,
the ground state has fourfold degeneracy.
We note that the local spin is quenched completely for $\tilde{\Delta}=0$.

\begin{figure}[t]
\begin{minipage}{8cm}
\epsfxsize=8cm
\epsfbox{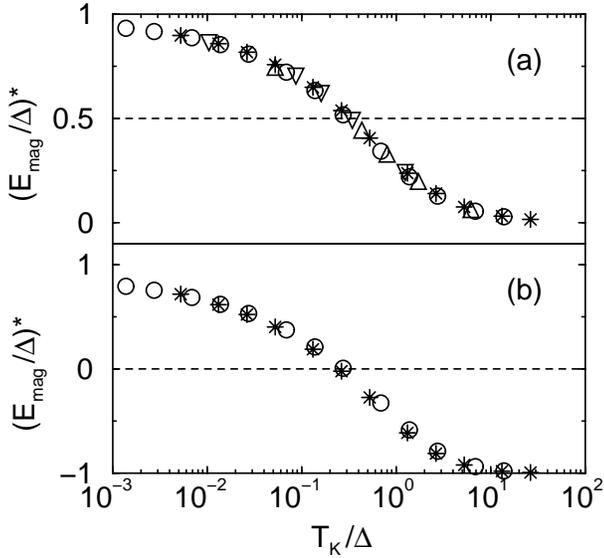}
\end{minipage}
\caption{
$T_{\rm K}/\Delta$ dependence of the bound state energy $(E_{\rm mag}/\Delta)^*$
generated by the magnetic impurity.
$(E_{\rm mag}/\Delta)^*$ is measured from the lowest-lying doublet
for each $T_{\rm K}/\Delta$.
(a) $p_x +i p_y$-wave case.
The ground state is a spin doublet
and the first excited state is a particle-hole doublet with no spin (see Table I).
The results for fixed $T_{\rm K}$ are expressed by a circle and a star
for $T_{\rm K}=9.20\times 10^{-5}$ and $1.76\times 10^{-3}$, respectively.
Triangle-up and triangle-down are the results for fixed $\tilde{\Delta}=0.001$ and $0.005$,
respectively.
(b) $s$-wave case.
In the positive $(E_{\rm mag}/\Delta)^*$ region,
the ground state is a spin doublet and the first excited state is a spin singlet.
In the negative $(E_{\rm mag}/\Delta)^*$ region,
the ground and excited states are interchanged.
The level crossover takes place at around $T_{\rm K}/\Delta=0.3$.
\protect\cite{Satori}
}
\label{fig:2}
\end{figure}
When $\tilde{J}$ turns on for a finite $\tilde{\Delta}$,
a bound state appears below the energy gap due to the magnetic impurity.
Figure \ref{fig:2}(a) shows the $T_{\rm K}/\Delta$ dependence of the bound state energy
$(E_{\rm mag}/\Delta)^*$ for the $p_x +i p_y$-wave.
Here, the Kondo temperature is defined by
\begin{equation}
T_{\rm K}=\sqrt{|J|}{\rm exp}(-1/|J|).
\end{equation}
First, we look into the small $T_{\rm K}/\Delta$ region.
As shown in Table I(A), the ground state is a spin doublet.
The first excited state is a particle-hole doublet with no spin.
In the Kondo effect, it is favorable for the magnetic impurity to form a Kondo singlet.
However, the magnetic impurity has to break the Cooper pair
to couple with one of the paired electrons.
Since this costs considerable energy,
the Kondo singlet cannot be a ground state in the small $T_{\rm K}/\Delta$ region.
As $T_{\rm K}/\Delta$ increases,
$(E_{\rm mag}/\Delta)^*$ decreases monotonically
and approaches the ground state energy asymptotically.
Thus, the ground state approaches the $\tilde{\Delta}=0$ result smoothly
as $T_{\rm K}/\Delta$ increases to infinity,
and it becomes fourfold degenerate.
This result shows that the spin singlet ground state is not realized
in all the $T_{\rm K}/\Delta$ region,
which is completely different from the $s$-wave result shown in Fig. \ref{fig:2}(b).
On the other hand,
we find that $(E_{\rm mag}/\Delta)^*$ for the $p_x +i p_y$-wave is scaled by $T_{\rm K}/\Delta$
as for the $s$-wave.
While the interchange of the ground state does not occur in the $p_x +i p_y$-wave case,
almost the same $T_{\rm K}/\Delta$ dependence of $(E_{\rm mag}/\Delta)^*$ is obtained
in both cases.
For the $s$-wave, the energy gap suppresses the Kondo effect
in the $(E_{\rm mag}/\Delta)^*>0$ region,
while the Kondo singlet is stabilized against the energy gap
in the $(E_{\rm mag}/\Delta)^*<0$ region.
This crossover occurs at approximately $T_{\rm K}/\Delta=0.3$.
\cite{Satori}
In analogy to the $s$-wave, there is also a crossover at around $(E_{\rm mag}/\Delta)^*=0.5$
for the $p_x +i p_y$-wave.
Thus, the Kondo effect overcomes the energy gap in the $(E_{\rm mag}/\Delta)^*<0.5$ region,
although the ground state is still a spin doublet.
This implies that the competition between the Kondo effect and the energy gap is characterized by
such $T_{\rm K}/\Delta$ dependence of the bound state energy for all types of superconductivity.
We mention here that the NRG energy level structure for the $d_{x^2-y^2}+id_{xy}$-wave
is the same as that for the $p_x +i p_y$-wave.
Their difference appears only in the wave functions.
\begin{table}[t]
\caption{
Ground and first excited states.
(A) and (B) are the results for $T_{\rm K}/\Delta=2.76\times 10^{-2}$ and 1.38, respectively.
The energy $E$ is measured from the ground state (see Fig. \ref{fig:2}).
$Q$ represents the total quasiparticle number measured from the half-filling.
$S$ is the size of the total spin.
($p$) and ($s$) are the results for the $p_x +i p_y$-wave and $s$-wave, respectively.
For the $p_x +i p_y$-wave, the ground state is a spin doublet
and the first excited state is a particle-hole doublet with no spin.
For the $s$-wave, the results at the even number of renormalization step are given.
}
\begin{displaymath}
\begin{array}{crcccrcc} \hline \hline
 & (A) & & & & (B) & & \\ \hline
 & (E / \Delta)^* & Q & 2S & & (E / \Delta)^* & Q & 2S \\ \hline
(p)~~~ & 0.000 & 0 & 1 & ~~~~~~ & 0.000 & 0 & 1 \\
 & 0.808 & \pm 1 & 0 & ~~~~~~ & 0.220 & \pm 1 & 0 \\ \hline
(s)~~~ & 0.000 & 1 & 1 & ~~~~~~ & 0.000 & 0 & 0 \\
 & 0.530 & 0 & 0 & ~~~~~~ & 0.583 & 1 & 1 \\ \hline \hline
\end{array}
\end{displaymath}
\end{table}

Let us discuss the strong coupling limit ($|\tilde{J}|\rightarrow \infty$) case
where the impurity part $H_0$ is truncated from the conduction electron part.
In this limit, the ground state property is given by $H_0$.
Converting from the channel ($\tau$) representation to the angular momentum ($l$) representation
\begin{equation}
c_{N=0,\tau=\pm,\sigma}= \frac{1}{\sqrt{2}} (f_{l=0,\sigma} \pm f_{l=1,\sigma}),
\end{equation}
we can express $H_0=H_{s{\rm -}d}+H_\Delta$, where
\begin{eqnarray}
H_{s{\rm -}d} &=& -\tilde{J} \sum_{\sigma\sigma'}
  \mbox{\boldmath$S$} \cdot \mbox{\boldmath$\sigma$}_{\sigma\sigma'}
  f_{0\sigma}^\dagger f_{0\sigma'}, \cr
H_\Delta &=& - \tilde{\Delta} \sum_\sigma
  ( f_{0\sigma}^\dagger f_{1\sigma} + f_{1\sigma}^\dagger f_{0\sigma} ).
\label{eqn:limit}
\end{eqnarray}
Note that the order parameter corresponds to the hopping parameter
between the $l=0$ and $l=1$ local orbital sites, since
the $p_x +i p_y$-wave Cooper pair is formed by the $l=0$ and $l=1$ particles.
When $|\tilde{J}| \rightarrow \infty$,
the $l=0$ particle couples with the local spin strongly to form a spin singlet
$|s \rangle = f_{0\uparrow}^\dagger |\downarrow \rangle
             -f_{0\downarrow}^\dagger |\uparrow \rangle$.
Here, $|\sigma\rangle$ ($\sigma=\uparrow,\downarrow$) is the wavefunction of the local spin.
The spin doublet
$\left( f_{1\sigma}^\dagger |s \rangle \right)$
and spin singlet
$\left( f_{1\uparrow}^\dagger f_{1\downarrow}^\dagger |s \rangle,~|s \rangle \right)$
are degenerate since the hopping between $l=0$ and $l=1$ is forbidden.
At a finite $|\tilde{J}|$, $\tilde{\Delta}$ in eq. (\ref{eqn:limit}) lifts the degeneracy.
The wavefunctions for the spin doublet ($\psi_{S=1/2}$) and spin singlet ($\psi_{S=0}$)
are given by the linear combination of the following terms:
\begin{eqnarray}
\psi_{S=1/2} &=&
  c_1  \left( f_{1\sigma}^\dagger |s\rangle \right)
+ c_2  \left( f_{0\uparrow}^\dagger f_{0\downarrow}^\dagger |\sigma\rangle \right)
+ c_3  \left( f_{1\uparrow}^\dagger f_{1\downarrow}^\dagger |\sigma\rangle \right) \cr
&+&
  c_4  \left( f_{1,-\sigma}^\dagger f_{0\sigma}^\dagger |\sigma\rangle \right)
+ c_5  \left[ f_{1\sigma}^\dagger \left( f_{0\uparrow}^\dagger |\downarrow\rangle
                                  + f_{0\downarrow}^\dagger |\uparrow\rangle \right) \right], \cr
\psi_{S=0} &=&
  c_6 |s\rangle
+ c_7  \left( f_{1\uparrow}^\dagger |\downarrow\rangle
            - f_{1\downarrow}^\dagger |\uparrow\rangle \right).
\end{eqnarray}
Here, $c_i$ is a coefficient.
The perturbation study shows that $\psi_{S=1/2}$ is the ground state,
whose energy is lower than that of $\psi_{S=0}$ by
$-\frac{2}{3} |\tilde{J}| (\tilde{\Delta}/\tilde{J})^4$.
When $|\tilde{J}|$ is very large,
coefficient $c_1$ of the first term in $\psi_{S=1/2}$ is the largest.
In the strong coupling limit, the local spin is almost quenched by the $l=0$ particle
forming the Kondo singlet $|s\rangle$.
However, the $l=1$ particle is connected weakly to the quenched local spin $|s\rangle$.
This does not mean that the $l=1$ electrons destroy the Kondo singlet.
They gain the superconducting condensation energy
and then generate the spin 1/2 of the ground state in the strong coupling limit.
As $|\tilde{J}|$ decreases,
coefficients $c_2$ and $c_3$ of the second and third terms,
respectively,
in $\psi_{S=1/2}$ increase.
This indicates that the weight of the local spin becomes larger in the spin of the ground state.
Thus, the ground state always has a spin 1/2 for the $p_x +i p_y$-wave.
The $T_{\rm K}/\Delta$ dependence of the bound state energy shown in Fig. \ref{fig:2}(a)
reflects the continuous change of the wavefunction of the spin doublet ground state.

Contrary to the $p_x +i p_y$-wave,
the wavefunctions for the $s$-wave are given by $\varphi_{S=0}=|s\rangle$
and $\varphi_{S=1/2}=f_{0\uparrow}^\dagger f_{0\downarrow}^\dagger |\sigma\rangle$
in the strong coupling limit.
In this case, the hopping between the $l=0$ and $l=1$ orbital sites is replaced
by the potential at the $l=0$ site:
\begin{equation}
H_\Delta = - \tilde{\Delta}
( f_{0\uparrow}^\dagger f_{0\uparrow} + f_{0\downarrow}^\dagger f_{0\downarrow} -1 ).
\end{equation}
This is because the $s$-wave Cooper pair is formed by only the $l=0$ particles.
The ground state is characterized by $\varphi_{S=0}$ ($\varphi_{S=1/2}$)
in a large (small) $T_{\rm K}/\Delta$ region.
For the $s$-wave,
the $l\ne 0$ conduction electrons are completely decoupled from the local spin.

In conclusion, we have studied the Kondo effect unique to the $p_x +i p_y$-wave superconductor,
where the orbital dynamics of the Cooper pair produces the spin of the ground state.
For large $T_{\rm K}$,
the $l=0$ electrons couple with a local spin strongly to form a Kondo singlet,
while the other electrons with $l=1$ are coupled to the Kondo singlet weakly
via the superconducting order parameter.
The NRG study has shown that the ground state is a spin doublet
in all the $T_{\rm K}/\Delta$ region,
which is different from the results for the $s$-wave and $d_{x^2-y^2}$-wave cases.
It is interesting to search for a new type of Kondo effect
due to the orbital effect of the Cooper pair in unconventional superconductors.

We are grateful to H. Kusunose, R. Shiina, and K. Ueda for helpful comments.



\end{document}